\documentclass[aps,prc,preprint,amsmath,amssymb,showpacs,preprintnumbers,superscriptaddress]{revtex4-1}
\usepackage{CJK}
\usepackage{graphicx}
\usepackage{dcolumn}
\usepackage{bm}
\usepackage{color}
\usepackage{amsmath}
\allowdisplaybreaks[4]

\usepackage{hyperref}

\begin{document}

\title{Pseudospin symmetry and octupole correlations for M$\chi$D in $^{131}\mathrm{Ba}$}

\author{Y. P. Wang}
\affiliation{State Key Laboratory of Nuclear Physics and Technology, School of Physics, Peking University, Beijing 100871, China}
\author{Y. Y. Wang}
\affiliation{State Key Laboratory of Nuclear Physics and Technology, School of Physics, Peking University, Beijing 100871, China}
\author{J. Meng}
\affiliation{State Key Laboratory of Nuclear Physics and Technology, School of Physics, Peking University, Beijing 100871, China}
\affiliation{Yukawa Institute for Theoretical Physics, Kyoto University, Kyoto 606-8502, Japan}

\date{\today}
\begin{abstract} 

A Reflection-Asymmetric Triaxial Particle Rotor Model (RAT-PRM) with three quasiparticles and a reflection-asymmetric triaxial rotor is developed, and applied to investigate the observed multiple chiral doublets (M$\chi$D) candidates with octupole correlations in $^{131}$Ba, i.e., two pairs of positive-parity bands D3-D4 and D5-D6, as well as one pair of negative-parity bands D7-D8.
The energy spectra, the energy staggering parameters, the $B(M1)/B(E2)$ ratios, and the $B(E1)/B(E2)$ ratios are reproduced well. The chiral geometries for these M$\chi$D candidates are examined by the \emph{azimuthal plots}, and the evolution of chiral geometry with spin is
clearly demonstrated. The intrinsic structure for the positive-parity bands is analyzed and the possible pseudospin-chiral quartet bands are suggested.

\end{abstract}
\maketitle

\date{today}

\section{Introduction}\label{sec1}

Chirality is a phenomenon of general interest in chemistry, biology and particle physics. 
Since the concept of chirality in atomic nuclei was first proposed by Frauendorf and Meng~\cite{Frauendorf1997Tilted}, many efforts have been made to understand chiral symmetry and its spontaneous breaking in atomic nuclei, see e.g., reviews~\cite{Frauendorf2001Spontaneous,Meng2010Open,Meng2016Nuclear,
Starosta2017Phys.Scripta93002,Frauendorf2018Phys.Scripta43003}.

Based on the adiabatic and configuration fixed constrained triaxial covariant density functional theory (CDFT)~\cite{meng2016relativistic}, a new phenomenon named multiple chiral doublets (M$\chi$D) is predicted~\cite{Meng2006Possible}. The chiral doublet bands are a pair of nearly degenerate $\Delta I=1$ bands with the same parity. M$\chi$D suggest that more than one pair of chiral doublet bands can exist in one single nucleus. 
The first experimental evidence for M$\chi$D is reported in $^{133}$Ce~\cite{Ayangeakaa2013Evidence}, and 
followed by $^{103}$Rh \cite{Kuti2014Multiple}, 
$^{78}$Br~\cite{Liu2016Phys.Rev.Lett.112501}, $^{136}$Nd~\cite{Petrache2018Phys.Rev.C41304} 
and $^{195}$Tl~\cite{Roy2018Phys.Lett.768}, etc. 
More than 60 chiral doublet candidates in around 50 nuclei (including 10 nuclei with M$\chi$D) have been reported in the $A\sim$~80, 100, 130 and 190~mass regions~\cite{Xiong2019Atom.DataNucl.DataTabl.193,
Wang2018Phys.Rev.14304,Roy2018Phys.Lett.768,Guo2019PRL}.

Theoretically, nuclear chirality has been investigated with many approaches, including the triaxial particle rotor model (PRM)~\cite{Frauendorf1997Tilted,Peng2003Description,
Koike2004Chiral,Zhang2007Phys.Rev.C44307,Qi2009Chirality,Chen2018},
the tilted axis cranking (TAC) model~\cite{Frauendorf1997Tilted,Dimitrov2000Chirality,
Olbratowski2004Critical,Olbratowski2006Search,Zhao2017Phys.Lett.B1}, the TAC approach with the random phase approximation (TAC+RPA)~\cite{Mukhopadhyay2007From,Almehed2011Chiral} and the collective Hamiltonian (TAC+CH)~\cite{Chen2013Collective,Chen2016Two,Wu2018Phys.Rev.64302}, the interacting boson-fermion-fermion model (IBFFM)~\cite{Brant2008Phys.Rev.34301}, the generalized coherent state model \cite{Raduta2016J.Phys.95107}
and the projected shell model (PSM)~\cite{Hara1995PROJECTED,Bhat2014Investigation,Chen2017Chiral,Chen2018Phys.Lett.211,Wang2019PRC}.

The PRM is a quantal model coupling the collective rotation and the single-particle motions in the laboratory reference frame,
and describes directly the quantum tunneling and energy splitting between the doublet bands.
Various versions of PRM have been developed for the nuclear chirality.
For odd-odd nuclei, the triaxial PRM with 1-particle-1-hole configuration is mainly used~\cite{Frauendorf1997Tilted,Peng2003Description,Koike2004Chiral,Qi2009Examining}.
To simulate the effects of many valence nucleons, the triaxial PRM with 2-qusiparticle configuration was developed with pairing correlations taken into account by the Bardeen-Cooper-Schrieffer (BCS) approximation~\cite{Zhang2007Phys.Rev.C44307,Koike2003,Wang2007Examining,Wang2008Description,Lawrie2008Possible,Lawrie2010,Shirinda2012}.
For odd-A and even-even nuclei, the many-particle-many-hole versions of triaxial PRM with nucleons in 2$j$ shells~\cite{Qi2009Chirality,Qi2011Chirality}, 3$j$ shells~\cite{Ayangeakaa2013Evidence,Kuti2014Multiple,Lieder2014Resolution,Qi2013Possible}, and 4$j$ shells~\cite{Chen2018}, have been developed.
The recently observed octupole correlations between the M$\chi$D candidates in $^{78}$Br~\cite{Liu2016Phys.Rev.Lett.112501} stimulate the development of the reflection-asymmetric triaxial PRM (RAT-PRM) with both triaxial and octupole degrees of freedom~\cite{WANG2019454}.

The RAT-PRM in Ref.~\cite{WANG2019454} is based on a reflection-asymmetric triaxial rotor coupled with two quasiparticles.
In Ref.~\cite{Guo2019PRL}, the M$\chi$D candidates with octupole correlations, i.e., two pairs of positive-parity bands D3-D4 and D5-D6, as well as one pair of negative-parity bands D7-D8 are observed in nucleus $^{131}$Ba.
The positive-parity bands, D3-D4 and D5-D6 with $\pi h_{11/2}(g_{7/2},d_{5/2})\otimes \nu h_{11/2}$, and the negative-parity bands D7-D8 with $\pi h_{11/2}^2\otimes \nu h_{11/2}$, are built on 3-quasiparticle configurations.
Furthermore, the four nearly degenerate positive-parity bands are suggested to be the pseudospin-chiral quartet bands~\cite{Guo2019PRL}.

In this work, a RAT-PRM with three quasiparticles and a reflection-asymmetric triaxial rotor is developed, and applied to investigate the observed M$\chi$D with octupole correlations in $^{131}$Ba.
The model is introduced in Sec. \ref{Sec2}, and the numerical details are presented in Sec. \ref{Sec3}. The energy spectra, the electromagnetic transitions, and the angular momentum orientations are calculated and compared with the data available in Sec. \ref{Sec4}, and a summary is given in Sec. \ref{Sec5}.

\section{Formalism}\label{Sec2}

The total RAT-PRM Hamiltonian can be expressed as
\begin{equation}
  \hat { H } = \hat { H } _ { \mathrm { core } } + \hat { H } _ { \mathrm { intr. } }^\pi+ \hat { H } _ { \mathrm { intr. } }^\nu,
\end{equation}
where $\hat{H}_{\mathrm{core}}$ is the Hamiltonian of a reflection-asymmetric triaxial rotor and $\hat{H}_{\mathrm{intr.}}^{\pi(\nu)}$ is the intrinsic Hamiltonian for valence protons (neutrons) in a reflection-asymmetric triaxially deformed potential.

Similar as Ref.~\cite{WANG2019454}, the core Hamiltonian generalized straightforwardly from the reflection-asymmetric axial rotor~\cite{Leander1984Nucl.Phys.375} reads,

\begin{equation}
\hat { H } _ { \mathrm { core } }  = \sum _ { k = 1 } ^ { 3 } \frac { \hat { R } _ { k } ^ { 2 } } { 2 \mathcal { J } _ { k } } + \frac { 1 } { 2 } E \left( 0 ^ { - } \right) ( 1 - \hat { P }_{c} ),
\end{equation}
with $\hat{R_k}=\hat{I}_k-\hat{J}_k$, the indices $k=1, 2, 3$ refer to the three principal axes of the body-fixed frame, and $\hat{R}_k$, $\hat{I}_k$, and $\hat{J}_k$ denote the angular momentum operators for the core, the nucleus and the valence nucleons, respectively. The moments of inertia for irrotational flow, $\mathcal { J } _ { k }=\mathcal { J } _ { 0 }\sin^2(\gamma-2k\pi/3)$, are adopted. The core parity splitting $E(0^-)$ is viewed as the excitation energy of the virtual $0^-$ state~\cite{Leander1984Nucl.Phys.375}. The product of the core parity operator $\hat{P}_c$ and the proton (neutron) single-particle parity operators $\hat{p}_{\pi(\nu)}$ gives the total parity operator $\hat{P}$.

The intrinsic Hamiltonian for valence nucleons is~\cite{WANG2019454}
\begin{equation}\label{hsp}
\begin{split}
  \hat{H}_{\mathrm{intr.}}^{\pi(\nu)}&=\hat{H}_{\mathrm{s.p.}}^{\pi(\nu)}+\hat{H}_{\mathrm{pair}}\\
  & =\sum_{\tau>0}\left(\varepsilon_{\tau_{\pi(\nu)}}-\lambda_{\pi(\nu)}\right)\left(a_{\tau_{\pi(\nu)}}^{\dagger} a_{\tau_{\pi(\nu)}}+a_{\overline{\tau}_{\pi(\nu)}}^{\dagger} a_{\overline{\tau}_{\pi(\nu)}}\right)-\frac{\Delta}{2} \sum_{\tau>0}\left(a_{\tau_{\pi(\nu)}}^{\dagger} a_{\overline{\tau}_{\pi(\nu)}}^{\dagger}+a_{\overline{\tau}_{\pi(\nu)}} a_{\tau_{\pi(\nu)}}\right),
\end{split}
\end{equation}
where $\lambda_{\pi(\nu)}$ denotes the Fermi energy of proton (neutron), $\Delta$ is the pairing gap parameter, and the single-particle energy $\varepsilon_{\tau_{\pi(\nu)}}$ is obtained by diagonalizing the reflection-asymmetric triaxial Nilsson Hamiltonian $\hat{H}_{s.p.}^{\pi(\nu)}$~\cite{WANG2019454}. The indices $\tau_{\pi(\nu)}$ and $\overline{\tau}_{\pi(\nu)}$ represent the proton (neutron) single-particle state and its corresponding time reversal state, respectively.

Taking into account the pairing correlations by the BCS approach~\cite{Hamamoto1976Nucl.Phys.15,Hamamoto1983Phys.Lett.281,Zhang2007Phys.Rev.C44307,WANG2019454}, the intrinsic Hamiltonian becomes
\begin{equation}
\hat{H}_{\mathrm{intr.}}^{\pi(\nu)}=\sum_{\tau_{\pi(\nu)}>0} \varepsilon_{\tau_{\pi(\nu)}}^{\prime}\left(\alpha_{\tau_{\pi(\nu)}}^{\dagger} \alpha_{\tau_{\pi(\nu)}}+\alpha_{\overline{\tau}_{\pi(\nu)}}^{\dagger} \alpha_{\overline{\tau}_{\pi(\nu)}}\right),
\end{equation}
with the quasiparticle energies $\varepsilon_{\tau_{\pi(\nu)}}^{\prime}=\sqrt{\left(\varepsilon_{\tau_{\pi(\nu)}}-\lambda_{\pi(\nu)}\right)^{2}+\Delta^{2}}$. 

In reference \cite{WANG2019454}, the RAT-PRM with a quasi-proton and a quasi-neutron coupled with a reflection-asymmetric triaxial rotor is developed. In order to describe the M$\chi$D bands with more than two quasiparticles, the intrinsic wavefunction for a system with $Z$ valence protons and $N$ valence neutrons in RAT-PRM is generalized as,
\begin{equation}\label{eq3}
  {\chi=\left(\prod_{i=1}^{Z}\alpha_{\tau_{\pi i}}^\dagger\right)\left(\prod_{j=1}^{N}\alpha_{\tau_{\nu j}}^\dagger\right)|0\rangle.}
\end{equation}
For $^{131}\mathrm{Ba}$, $Z=2$ and $N=1$.

The total Hamiltonian $\hat{H}$ is invariant under the operation $\hat{S}_2=\hat{P}_ce^{i\pi\hat{R}_2}$, which is the reflection with respect to the plane perpendicular to 2-axis. It can be diagonalized in the symmetrized strong-coupled basis with good parity and angular momentum~\cite{WANG2019454},
\begin{equation}\label{eq6}
  {|\Psi _ { I M K \pm }\rangle = \frac{1}{2\sqrt{1+\delta_{K0}\delta_{\chi,\bar{\chi}}}} \left( 1 + \hat { S } _ { 2 } \right) | I M K \rangle \psi _ { \pm },}
\end{equation}
where $|IMK\rangle=\sqrt{\frac{2I+1}{8\pi^2}}D_{MK}^{I*}$ is the Wigner function, $\bar{\chi}=\hat{S}_2\chi$, and $\psi _ { \pm }$ are the intrinsic wavefunctions with good parity, 
\begin{align}
&\psi _ { + } = ( 1 + \hat { P } ) \chi\Phi _ { a }=( 1 + \hat{P}_c\hat{p}_\pi\hat{p}_\nu ) \chi\Phi _ { a },\label{eq4}\\
&\psi _ { - } = ( 1 - \hat { P } ) \chi\Phi _ { a }=( 1 - \hat{P}_c\hat{p}_\pi\hat{p}_\nu ) \chi\Phi _ { a }.\label{eq5}
\end{align}
Here, $\Phi _ { a }$ describes the same orientation in space between the core and the intrinsic single-particle potential, and $\chi\Phi_{a}$ is the core-coupled intrinsic wavefunction.

The diagonalization of the total Hamiltonian $\hat{H}$ gives the eigenvalues and
eigenfuctions. Then one can calculate the reduced electromagnetic transition probabilities~\cite{WANG2019454}, angular momentum components~\cite{WANG2019454}, and the probability profile for the orientation of the angular momentum, i.e., \emph{azimuthal plot}~\cite{Chen2017Chiral, Chen2018Phys.Rev.31303}, etc.
 
\section{Numerical details}\label{Sec3}

For the M$\chi$D candidates with octupole correlations observed in nuclei $^{131}$Ba, two pairs of positive-parity bands D3-D4 and D5-D6 are suggested to have the configuration $\pi h_{11/2}(g_{7/2},d_{5/2})\otimes \nu h_{11/2}$~\cite{Guo2019PRL}.
For this configuration, the microscopic configuration-fixed triaxial covariant density functional theory~\cite{Meng2006Possible}  with PC-PK1~\cite{Zhao2010Phys.Rev.C54319} gives the quadrupole deformation $\beta_2=0.22$ and $\gamma=27.1^{\circ}$. {An octupole deformation $\beta_3 = 0.02$ is adopted to investigate the octupole correlations in the present RAT-PRM calculations.}

With the deformation parameters above, the reflection-asymmetric triaxial Nilsson Hamiltonian is solved by expanding the wavefunction by harmonic oscillator basis~\cite{Wang2018Sci.ChinaPhys.Mech.Astron.82012}. 
According to the configuration suggested in Ref.~\cite{Guo2019PRL}, the Fermi energies of proton and neutron are chosen as $\lambda_\pi=43.86~\mathrm{MeV}$ and $\lambda_\nu=48.76~\mathrm{MeV}$, respectively. The single-particle space includes seven above and six below the Fermi level. The pairing correlation is taken into account by the empirical formula $\Delta=12/\sqrt{A}$ MeV.

The moment of inertia $\mathcal { J } _ { 0 }=19~\mathrm{\hbar^2/MeV}$ and the core parity splitting parameter $E(0^-)=2.1$ MeV are adjusted to the experimental spectra. For the calculations of the magnetic transitions, the gyromagnetic factors are $g_R=Z/A$ for the core and  $g_{\pi(\nu)}=g_l+(g_s-g_l)/(2l+1)$ for the proton (neutron)~\cite{bohr1975nuclear,ring2004nuclear}. For the calculations of the electric transitions, the empirical intrinsic dipole moment $Q_{10} = \frac{3}{4\pi}R_0Ze\beta_{10}$ and quadrupole moment $Q_0 = \frac{3}{\sqrt{5\pi}}R_{0}^2Ze\beta_2$ are taken with $R_0 = 1.2A^{1/3}$ fm~\cite{WANG2019454}.

\section{Results and discussion}\label{Sec4}

\begin{figure}[t]
  \centering
\includegraphics[width=0.9\linewidth]{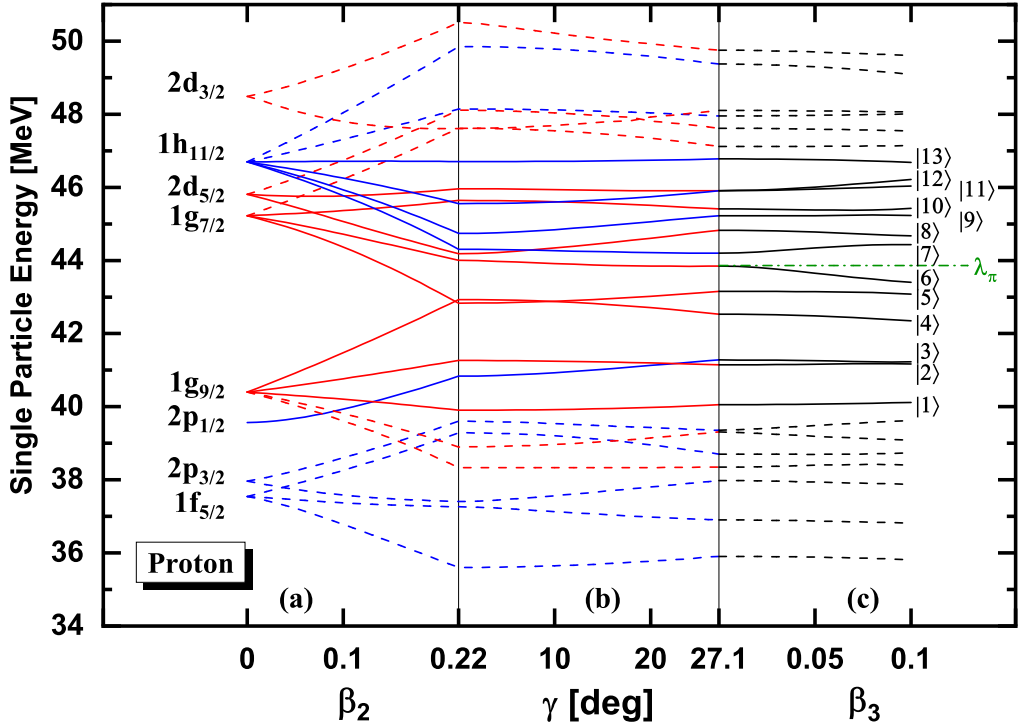}
\caption{The proton single-particle levels, obtained by diagonalizing the reflection-asymmetric triaxial Nilsson Hamiltonian in Eq.~(\ref{hsp}), as functions of the deformation parameters (a) $\beta_2$ ($\gamma=0^{\circ}$, $\beta_3=0$), (b) $\gamma$ ($\beta_2=0.22$, $\beta_3=0$), and (c) $\beta_3$ ($\beta_2=0.22$, $\gamma=27.1^{\circ}$).  The red and blue represent the positive and negative parity. The solid lines represent the chosen proton single-particle space in the RAT-PRM calculation. The green dash dot line represents the proton Fermi energy.}
  \label{fig1}
\end{figure}

By diagonalizing the reflection-asymmetric triaxial Nilsson Hamiltonian in Eq. (\ref{hsp}), one can obtain the proton (neutron) single-particle levels. 
As an example, Fig.~\ref{fig1} shows the proton single-particle levels as functions of the deformation parameters. 
The proton single-particle space, represented by the solid lines in Fig.~\ref{fig1}, includes seven above and six below the Fermi level. The quasiparticle states are obtained from the single-particle states at the deformation $\beta_2=0.22, \gamma=27.1^{\circ}$ and $\beta_3=0.02$ by the BCS approximation. From these quasiparticle states, the intrinsic wavefunctions with good parity can be constructed from Eqs. (\ref{eq3}), (\ref{eq4}) and (\ref{eq5}). Then the intrinsic wavefunctions with good parity are used to generate the symmetrized strong-coupled basis in Eq. (\ref{eq6}). By diagonalizing the total RAT-PRM Hamiltonian on the strong-coupled basis, the energies and eigenfunctions at each spin $I$ are obtained.

\begin{figure}[t]
  \centering
\includegraphics[width=0.85\linewidth]{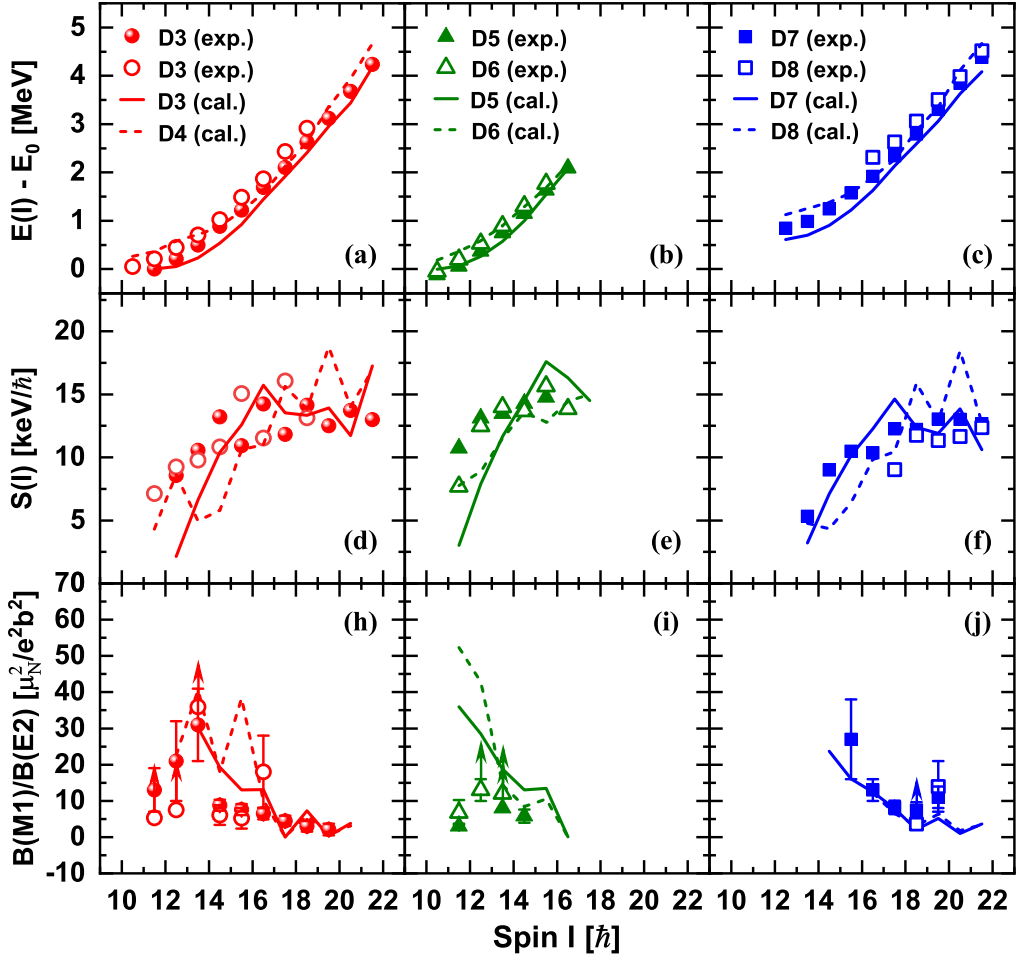}
\caption{The energies $E(I)$, the energy staggering parameters $S(I)=[E(I)-E(I-1)]/2I$, and the $B(M1)/B(E2)$ ratios for the two pairs of positive-parity bands D3-D4 (left panels) and D5-D6 (middle panels) as well as for the negative-parity doublet bands D7-D8 (right panels) in $^{131}$Ba by RAT-PRM (lines) in comparison with the experimental data (symbols) \cite{Guo2019PRL}. The bandhead energy of band D3 at $I=23/2\hbar$ is taken as a reference.}
  \label{fig2}
\end{figure}

In Fig.~\ref{fig2}, the energies $E(I)$, the energy staggering parameters $S(I)=[E(I)-E(I-1)]/2I$, and the $B(M1)/B(E2)$ ratios calculated by the RAT-PRM for the two pairs of positive-parity doublet bands D3-D4 and D5-D6 as well as for the negative-parity doublet bands D7-D8 are shown in comparison with the data available \cite{Guo2019PRL}. 
The bandhead of band D3 is taken as reference. 
The experimental spectra, the energy staggering parameters $S(I)$, and the $B(M1)/B(E2)$ ratios for bands D3-D8 are reproduced well by the RAT-PRM calculations. 
For each pair of the doublet bands, the spectra are nearly degenerate and the correspoding $B(M1)/B(E2)$ ratios  are similar, which are the characteristics for the chiral rotation~\cite{Wang2007Examining}.

\begin{figure}[t]
  \centering
\includegraphics[width=0.9\linewidth]{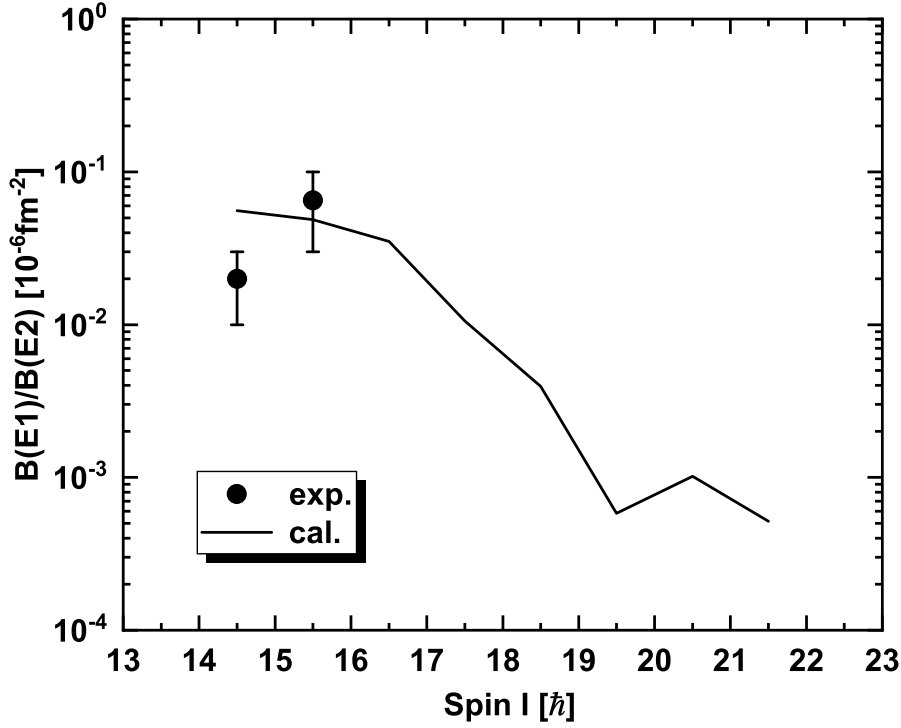}
\caption{The calculated $B(E1)/B(E2)$ ratios between the interband $E1$ transitions (band D7 $\rightarrow$ D3) and the intraband $E2$ transitions (band D7) in comparison with the data available~\cite{Guo2019PRL}.}
  \label{fig6}
\end{figure}

Since the octupole degree of freedom is included in the present RAT-PRM calculations, the electric dipole transition probabilities $B(E1)$ between the positive- and negative-parity bands can be calculated. 
In Fig.~\ref{fig6}, the calculated $B(E1)/B(E2)$ ratios between the interband $E1$ transitions (band D7 $\rightarrow$ D3) and the intraband $E2$ transitions (band D7) are shown in comparison with the data available~\cite{Guo2019PRL}. The trend of $B(E1)/B(E2)$ ratios are in general determined by that of the $B(E1)$ values. The experimental data are available at spin $I=29/2\hbar$ and $31/2\hbar$~\cite{Guo2019PRL} and can be well reproduced. Here the value of $\beta_3$ is chosen to reproduce the experimental $B(E1)/B(E2)$ data. Changing the $\beta_3$ by $10\%$, the influences on the energies, the staggering parameters, the $B(M1)/B(E2)$ ratios, and the chiral geometry are negligible, while the $B(E1)$ values are changed by about $20\%$.

\begin{figure}[t]
  \centering
\includegraphics[width=0.9\linewidth]{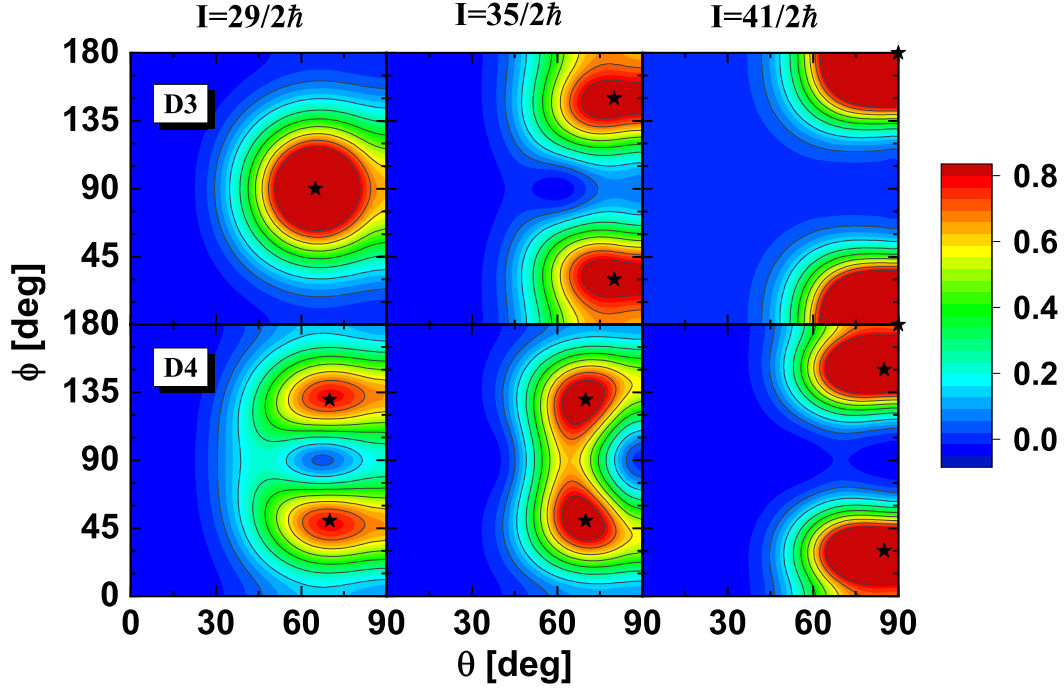}
\caption{The \emph{azimuthal plots}, i.e., probability distribution profiles for the orientation of the angular momentum on the $(\theta,\phi)$ plane, calculated for bands D3 and D4 at $I=29/2\hbar$, $35/2\hbar$, and $41/2\hbar$.  The black star represents the position of a local maximum.}
  \label{fig3}
\end{figure}

\begin{figure}[t]
  \centering
\includegraphics[width=0.9\linewidth]{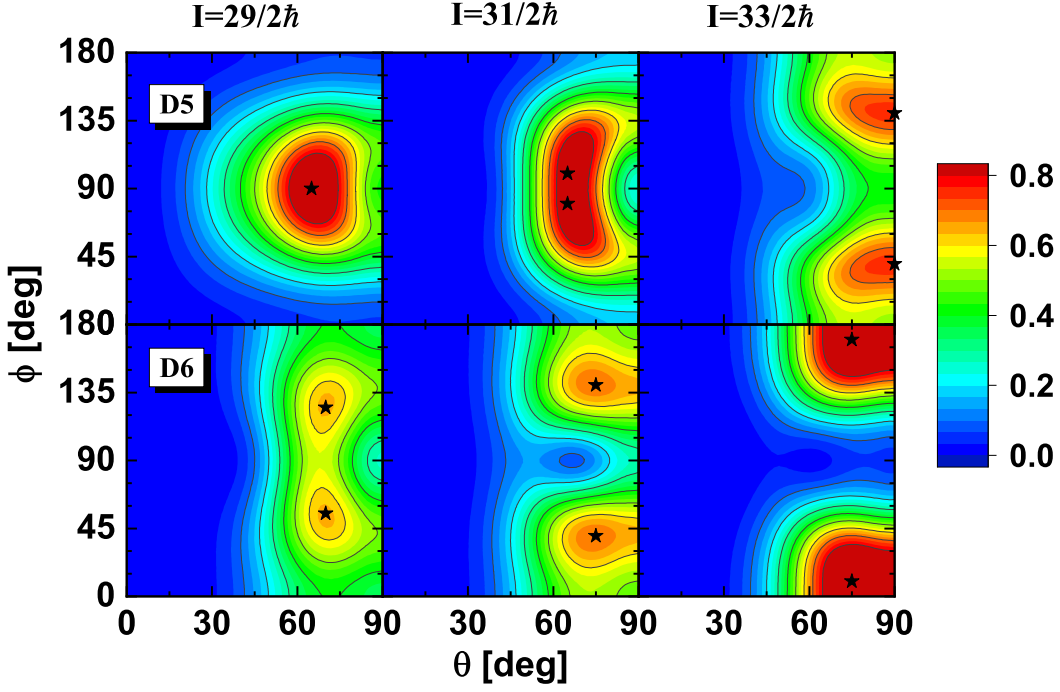}
\caption{Same as Fig. \ref{fig3}, but calculated for bands D5 and D6 at $29/2\hbar$, $31/2\hbar$, and $33/2\hbar$.}
  \label{fig4}
\end{figure}

\begin{figure}[t]
  \centering
\includegraphics[width=0.9\linewidth]{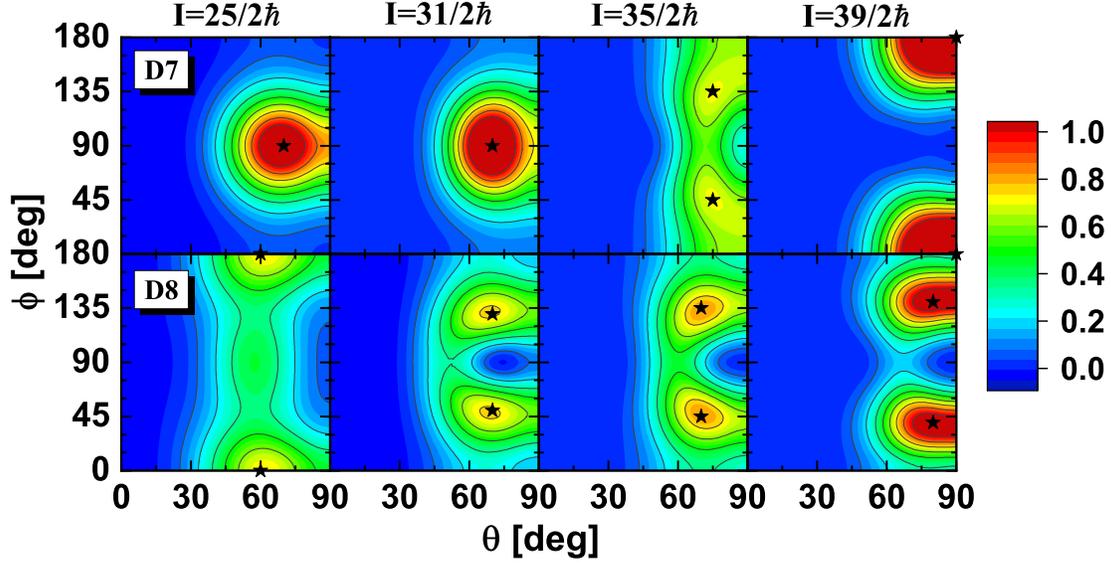}
\caption{Same as Fig. \ref{fig3}, but calculated for bands D7 and D8 at $I=25/2\hbar$, $31/2\hbar$, $35/2\hbar$ and $39/2\hbar$.}
  \label{fig5}
\end{figure}

In order to investigate the chiral geometry for the three chiral doublet candidates, the \emph{azimuthal plots}~\cite{Chen2017Chiral}, i.e., the probability distribution profiles for the orientation of the angular momentum on the $(\theta,\phi)$ plane, are shown in Figs.~\ref{fig3}~-~\ref{fig5}. 
The polar angle $\theta$ is the angle between the angular momentum and the intrinsic 3-axis, and the azimuthal angle $\phi$ is the angle between the projection of the angular momentum on the intrinsic 1-2 plane and the intrinsic 1-axis.
For $\gamma=27.1^\circ$, the 1, 2, and 3 axes are respectively the intermediate ($i$), short ($s$), and long ($l$) axes.

The \emph{azimuthal plots} for bands D3-D4 are shown in Fig.~\ref{fig3}. At spin $I=29/2\hbar$, the \emph{azimuthal plot} for band D3 has a single peak at $(65^{\circ}, 90^{\circ})$, which suggests that the angular momentum stays within
the $s$-$l$ plane, in accordance with the expectation for a 0-phonon state.
The \emph{azimuthal plot} for band D4 shows a node at $(65^{\circ}, 90^{\circ})$, and two peaks at $(70^{\circ}, 50^{\circ})$ and $(70^{\circ}, 130^{\circ})$. The existence of the node and the two peaks supports that the state in band D4 can be interpreted as a 1-phonon chiral vibration across the $s$-$l$ plane on the state in band D3.

At spin $I=35/2\hbar$, the \emph{azimuthal plots} for bands D3 and D4 are similar, with two peaks at $(80^{\circ}, 30^{\circ})$ and $(80^{\circ}, 150^{\circ})$ for band D3, and $(70^{\circ}, 50^{\circ})$ and $(70^{\circ}, 130^{\circ})$ for band D4. 
This shows the feature of static chirality.

At spin $I=41/2\hbar$, the peaks of the \emph{azimuthal plot} for band D3 are at $(90^{\circ}, 0^{\circ})$ and $(90^{\circ}, 180^{\circ})$, namely along the $i$ axis.
The peaks of the \emph{azimuthal plot} for band D4 locate $(85^{\circ}, 30^{\circ})$ and $(85^{\circ}, 150^{\circ})$, which are similar to those at $I=35/2\hbar$ but approaching the $i$-axis. The tendency from chiral rotation to principle axis rotation at high spins is shown.

In Fig.~\ref{fig4}, the \emph{azimuthal plots} for bands D5-D6 are shown. Similar as the evolution of the chiral geometry for bands D3-D4 in Fig.~\ref{fig3}, bands D5-D6 show the chiral vibration along the $\phi$ direction at $I=29/2\hbar$, static chirality at $I=31/2\hbar$, and a tendency to a princpal axis rotation around the $i$ axis at $I=33/2\hbar$. 

In Fig.~\ref{fig5}, the \emph{azimuthal plots} for negative-parity doublet bands D7-D8 are shown. At $I=25/2\hbar$, the peak of the \emph{azimuthal plot} for band D7 is at $(70^{\circ}, 90^{\circ})$, which is a planar rotation within the $s$-$l$ plane. The angular momentum is mainly contributed by the valence protons and the neutron hole. For band D8, the peaks are at $(60^{\circ}, 0^{\circ})$ and $(60^{\circ}, 180^{\circ})$, which corresponds to an $i$-$l$ planar rotation. The angular momentum is mainly contributed by the core and the valence neutron hole. The contribution of the valence protons are negligible. For $I\geq 33/2\hbar$, similar as the evolution of the chiral geometry for bands D3-D4 in Fig.~\ref{fig3}, bands D7-D8 show the chiral vibration along the $\phi$ direction at $I=31/2\hbar$, static chirality at $I=35/2\hbar$, and a tendency to a princpal axis rotation about the $i$ axis at $I=39/2\hbar$. 

The positive-parity M$\chi$D candidates observed in nuclei $^{131}$Ba, bands D3-D4 and D5-D6, are suggested to have the configuration $\pi h_{11/2}(g_{7/2},d_{5/2})\otimes \nu h_{11/2}$~\cite{Guo2019PRL}. One possible M$\chi$D mechanism is that the ``yrast'' and ``excited'' chiral doublets are built on the same configuration, similar as $^{103}$Rh in Ref.~\cite{Kuti2014Multiple}. Another possible mechanism is the appearance of the pseudospin-chiral quartet bands due to the pseudospin partners $(g_{7/2},d_{5/2})$.

\begin{figure}[t]
  \centering
\includegraphics[width=0.9\linewidth]{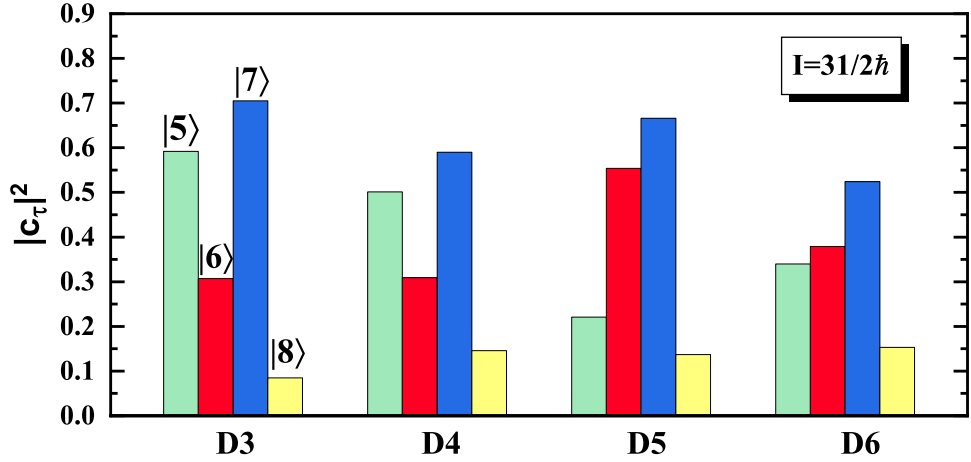}
\caption{The occupation of the single-particle level $|c_\tau|^2$ contributed from the corresponding strong-coupled basis which includes $|\tau\rangle$ in its intrinsic wavefunction with $\tau=5,6,7,8$ at $I=31/2\hbar$ for bands D3-D6.}
  \label{fig7}
\end{figure}

To pin down the mechanism for the positive-parity M$\chi$D candidates in $^{131}$Ba, the main occupation of the valence protons are investigated. In Fig.~\ref{fig7}, the occupation of the single-particle level $|c_\tau|^2$ is shown with $\tau=5,6,7,8$ at $I=31/2\hbar$ for bands D3-D6, which is calculated from the corresponding strong-coupled basis including $|\tau\rangle$ in its intrinsic wavefunction. The occupation patterns are similar for each chiral doublet bands. Examining more carefully, the top components for bands D3-D4 are levels $|5\rangle$ and $|7\rangle$, while for bands D5-D6 are levels $|6\rangle$ and $|7\rangle$.

\begin{figure}[t]
  \centering
\includegraphics[width=0.9\linewidth]{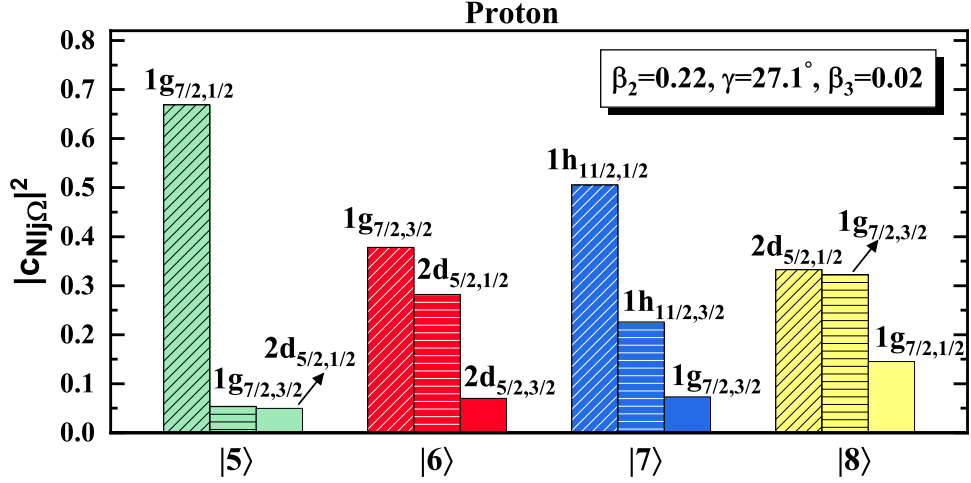}
\caption{The main components $Nlj\Omega$ of the proton single-particle levels $|\tau\rangle$ with $\tau=5,6,7,8$.}
  \label{fig9}
\end{figure}

As shown in Fig. \ref{fig1}, the levels $|5\rangle$ and $|6\rangle$ stem from $1g_{7/2}$, the level $|7\rangle$ stems from $1h_{11/2}$, and the level $|8\rangle$ stems from $2d_{5/2}$. Figure \ref{fig9} shows the main components $Nlj\Omega$ for the levels $|5\rangle$, $|6\rangle$, $|7\rangle$, and $|8\rangle$ in the present calculation with $\beta_2=0.22, \gamma=27.1^{\circ}, \beta_3=0.02$. The dominent components for the levels $|5\rangle$ and $|7\rangle$ are respectively $1g_{7/2}$ and $1h_{11/2}$. For the levels $|6\rangle$ and $|8\rangle$, there is a strong mixture of $1g_{7/2}$ and $2d_{5/2}$. Therefore the dominent component of the intrinsic wavefunctions is $\pi h_{11/2}g_{7/2}\otimes\nu h_{11/2}$ for bands D3-D4, in comparison with $\pi h_{11/2}(g_{7/2},d_{5/2})\otimes \nu h_{11/2}$ for bands D5-D6.

Thus the positive-parity M$\chi$D candidates involve the pseudospin partners $(g_{7/2},d_{5/2})$ and are suggested to be pseudospin-chiral quartet bands.

\section{Summary}\label{Sec5}

A Reflection-Asymmetric Triaxial Particle Rotor Model (RAT-PRM) for three quasiparticles coupled with a reflection-asymmetric triaxial rotor is developed, and applied to investigate the observed multiple chiral doublets (M$\chi$D) candidates with octupole correlations in $^{131}$Ba, i.e., two pairs of positive-parity bands D3-D4 and D5-D6, as well as one pair of negative-parity bands D7-D8.

The calculated energy spectra, energy staggering parameters $S(I)=[E(I)-E(I-1)]/2I$, $B(M1)/B(E2)$ ratios and $B(E1)/B(E2)$ ratios reproduce well the data available~\cite{Guo2019PRL}. The chiral geometries for the three pairs of chiral doublet bands are clearly illustrated by the \emph{azimuthal plots}. They show the chiral vibration at low spins, static chirality at intermediate spins, and a tendency to a princpal axis rotation around the $i$ axis at high spins. For band D8, an $i$-$l$ planar rotation occurs before the chiral vibration. The possible pseudospin-chiral quartet bands are suggested for the positive-parity M$\chi$D candidates D3-D6 based on their intrinsic structure.

\begin{acknowledgments}
 Fruitful discussions with S. Guo, C. M. Petrache, Y. K. Wang, S. Q. Zhang and P. W. Zhao are acknowledged.
 This work was partly supported by  the National Natural Science Foundation of China (Grants No. 11875075, No. 11935003, No. 11975031, and No. 11621131001), the National Key R\&D Program of China (Contracts No. 2018YFA0404400 and No. 2017YFE0116700), and the State Key Laboratory of Nuclear Physics and Technology, Peking University (No. NPT2020ZZ01).
\end{acknowledgments}

\end{document}